%% file: main.tex
\documentclass{article}
\usepackage{spconf,amsmath,amssymb,graphicx,booktabs,multirow,array,url,xcolor}
\usepackage{cite}
\usepackage{enumitem}
\usepackage[hidelinks]{hyperref}   % clickable cite/ref links; hidelinks = no coloured boxes

\makeatletter
\renewcommand{\thebibliography}[1]{\section{References}\footnotesize\list
 {[\arabic{enumi}]}{\settowidth\labelwidth{[#1]}\leftmargin\labelwidth
 \advance\leftmargin\labelsep
 \itemsep\z@ \parsep\z@ \topsep\z@ \partopsep\z@ \labelsep 0.4em
 \usecounter{enumi}}
 \def\newblock{\hskip .11em plus .33em minus .07em}
 \sloppy\clubpenalty4000\widowpenalty4000
 \sfcode`\.=1000\relax}
\makeatother

\newcommand{\method}{PromptSep-Speaker}
\newcommand{\promptsep}{PromptSep}

\newcommand{\sisdri}{\text{SI-SDRi}}

\newcommand{\tgt}{\mathrm{tgt}}

\title{PromptSep-Speaker: Generative Target Speaker Extraction via Multimodal Prompting and Semantic Alignment}
\name{\begin{tabular}{@{}c@{}}
Yutong Wen$^{1,2}$, Ke Chen$^{1}$, Prem Seetharaman$^{1}$, Haonan Chen$^{1}$,
Oriol Nieto$^{1}$\\
Paris Smaragdis$^{2}$, Justin Salamon$^{1}$, Zeyu Jin$^{1}$
\end{tabular}}
\address{$^{1}$Adobe Research, USA \qquad $^{2}$Massachusetts Institute of Technology, USA}

\begin{document}
\ninept
\maketitle

\begin{abstract}
\vspace{-0.15cm}
Target speaker extraction (TSE) isolates a desired speaker from a multi-speaker mixture using cues that identify the target. Prior work has explored acoustic and textual cues, such as enrollment audio and different types of speaker descriptors, but typically relies on a single cue family. Yet no single cue is both broadly available and reliably discriminative. Moreover, most TSE systems adopt mask-prediction models, which are prone to audible artifacts and generalize poorly across acoustic conditions. To address these limitations, we propose \method{}, a generative TSE model based on latent diffusion that unifies four interchangeable cues: speaker temporal order, transcript, enrollment audio, and speaker caption. The generative formulation delivers higher separation quality and multimodal prompting improves applicability across scenarios. To fully exploit each cue, we further introduce a frozen WavLM representation with an auxiliary alignment loss that supervises query-mixture cross-attention without extra labeling. Across four evaluation sets and a listening study, \method{} achieves state-of-the-art performance over TSE and universal source separation baselines, improving separation accuracy and quality across cue modalities.

\end{abstract}

\vspace{-0.2cm}
\section{Introduction}
\label{sec:intro}
\vspace{-0.3cm}
Speech mixtures with multiple speakers are prevalent in meetings, interviews, podcasts, and films. Rather than blindly separating speakers~\cite{hershey2016deep,luo2019convtasnet}, target speaker extraction (TSE) conditions separation on a cue that identifies the desired speaker and returns it alone~\cite{zmolikova2019sppeakerbeam,wang2019voicefilter}.

Early TSE systems adopt enrollment audio, conditioning separation on a voice clip of the target~\cite{zmolikova2019sppeakerbeam,wang2019voicefilter,xu2020spex,delcroix2020tdspeakerbeam}. This cue primarily refers to voice traits, however, making it prone to confusion when speakers sound alike or the clip is too short to convey distinctive traits~\cite{zhao2022target}. Textual cues have been explored as an alternative: LLM-TSE specifies the target in natural language~\cite{hao2023llmtse}, and subsequent work enriches this cue with LLM-provided context~\cite{mahmud2024opensep}, multiple speaker descriptors~\cite{huo2025styletse}, and relative attributes to another speaker~\cite{dai2025relative}. Such cues remove the need for enrollment audio, but typically identify a \emph{class} rather than an \emph{instance}: multiple speakers may satisfy the same description in some natural conversations. To pinpoint a specific speaker rather than a class, a concurrent work uses spoken words by the target~\cite{li2026daetse}. Other systems instead combine cues in the text family, such as pairing voice description with spatial information~\cite{zhu2025ssdq}.

While TSE has been studied extensively, most approaches only supports a single cue, and some multi-cue systems largely stay within one modality (usually text). Even those that combine audio and text in cues treat the two as complementary, such as text augmenting an enrollment audio~\cite{hao2023llmtse,huo2025styletse} rather than replacing it. These systems also rely primarily on mask-prediction models, which can be less robust in separation quality and across acoustic conditions than generative approaches~\cite{lemercier2023analysing}. Generative separation, by contrast, has achieved strong separation quality but mainly in universal sound source separation with free-form captions~\cite{yin2024flowsep,wang2025soloaudio,zhu2026audio, wen2026promptsep}; some speech-support models~\cite{samaudio} depends on visual cues and degrades under speech overlapping. Within TSE, generative approaches recently emerged and confined to enrollment audio conditioning~\cite{wang2024solospeech,hsieh2025adflowtse}. The most concurrent work, AuK~\cite{ma2026auk}, embeds separation as part of its foundation model but reports results on only one benchmark without per-cue breakdown. Additionally, evaluations in TSE are dominated by fully overlapped mixtures, but conversational speech interleaves partial overlap and turn-taking; models optimized for constant overlap may suppress segments on single-speaker region~\cite{oh2026vortex,realtse2026}.

We therefore propose \method{}, a generative latent-diffusion TSE separator that accepts any of four interchangeable cues: \textbf{when} they talk, \textbf{what} they say, \textbf{who} they sound like, and \textbf{how} they are described. Namely, a \textbf{speaker id} identifies the speaker by the
temporal order they begin speaking (e.g., \texttt{spk-1} speaks first)~\cite{fujita2019eend,bredin2020pyannote}; a \textbf{transcript} 
identifies the speaker through spoken content; an \textbf{enrollment} audio clip describes acoustic identity; and a \textbf{caption}
describes the speaker in free-form text. \method{} handles all four cues in a single model, so users supply whichever cue their setting affords rather than the one the system demands.

To better capture speech transcript and speaker order, we
incorporate a frozen speech representation,
WavLM~\cite{chen2022wavlm}, whose features are known to encode both phonetic and speaker
information and have been used as
auxiliary features for enhancement and separation~\cite{huang2022investigating}. However,
simply adding these features as another conditioning stream leaves a generative model
free to ignore them. We therefore propose a new \textbf{auxiliary alignment loss} that supervises how the model attends to the
cue, so that the speech representation is actively used to locate the target in the
mixture. Supervising attention in this way has precedent: guided attention aligns
text and audio in TTS~\cite{tachibana2018dctts}, and related objectives ground text to image
regions in diffusion models~\cite{chefer2023attendexcite,xie2023boxdiff}. We adapt the idea
to bind a speech cue to the target speaker. Our contributions are three-fold:
\begin{enumerate}[leftmargin=*,itemsep=1pt,parsep=0pt,topsep=1pt]
  \item \method{} a generative TSE model that accepts multimodal prompting: 
        enrollment audio, speaker id, transcript, and caption to identify the target, whereas prior work typically adopts mask-prediction models and studies one
        cue family at a time.
  \item We introduce a novel auxiliary alignment loss that supervises cue attention at no extra
        labeling cost, and show through ablations that it is the decisive component.
  \item We conduct a thorough evaluation across every cue on four evaluation sets and a listening test. \method{} reaches state-of-the-art results against both TSE and universal source separation baselines. A demo page is released for showcase\footnote{https://yutongwen.github.io/PromptSep-Speaker/}.
\end{enumerate}

\vspace{-0.3cm}
\section{Methodology}
\vspace{-0.2cm}

% =====================================================================
% \vspace{-0.1cm}
\subsection{Model Architecture}
% \vspace{-0.2cm}
Fig.~\ref{fig:arch} illustrates the architecture of \method{}, a conditional latent-diffusion model that maps a mixture waveform and
conditioning cues to the target source. The $48$~kHz mixture and enrollment audio are both encoded into latents by a frozen DAC-VAE~\cite{kumar2023dac,defossez2023encodec}; a
DiT module~\cite{peebles2023scalable} then generates the target latents, which the DAC-VAE decodes back to the separated waveform. 

Our DiT module follows the specifications of~\cite{garcia2025sketch2sound, wen2026promptsep}, and the text cues are encoded by Qwen3-VL-8B~\cite{bai2025qwen3}. Notably, the mixture is encoded by a
frozen WavLM base-plus model~\cite{chen2022wavlm} to provide semantic
features via an alignment module (in section 2.3 and 2.4).

\vspace{-0.3cm}
\subsection{Condition Injection}

\begin{figure}[t]
\centering
\includegraphics[width=\linewidth]{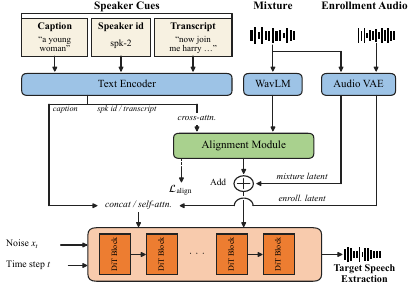}
\vspace{-0.7cm}
\caption{\method{} architecture.}
\label{fig:arch}
\vspace{-0.5cm}
\end{figure}

We consider four cues for target speaker extraction: 
\begin{itemize}[leftmargin=*,itemsep=1pt,parsep=0pt,topsep=1pt]
\item \textbf{caption}: a free-form description of the target speaker;
\item \textbf{speaker id}: a tag indicating the temporal order of each speaker by starting time (e.g., \texttt{spk-1}, \texttt{spk-2});
\item \textbf{transcript}: the words spoken by the target speaker;
\item \textbf{enrollment audio}: a reference clip of the target speaker speaking a different utterance, up to 8 seconds long.
\end{itemize}
% \subsection{Cue conditioning}
% \label{sec:routing}
% Each text prompt carries a special marker identifying its cue type (\texttt{<|caption|>}, \texttt{<|transcript|>}, or \texttt{<|spk$k$|>}), and a per-row splitter parses the prompt at the first cue marker and dispatches it accordingly:
As illustrated in Fig.~\ref{fig:arch}, the four cues take one of two routes through the model. Caption text tokens and enrollment audio latents are concatenated with the diffusion latent sequence, and the DiT module processes them jointly via self-attention. Speaker id and transcript text tokens instead enter the
alignment module, where they serves as cue tokens that the
WavLM representation of the mixture cross-attends to; then the resulting embedding is
added directly to the diffusion latent as condition.
% An alignment module serves both transcript and speaker-tag cues, with multi-cue rows containing both transcript and speaker information routed to the word-level objective. For enrollment, the reference clip is DAC-encoded and prepended to the latent sequence, and this slot is consumed only when enrollment audio is present, while text-only rows leave it masked. Finally, adaLN pooling of the text uses separate projections for the caption path and the grounded transcript/speaker path, preventing weight reuse across different cue types.
% Each text prompt carries a special marker that identifies its cue type
% (\texttt{<|caption|>}, \texttt{<|transcript|>}, \texttt{<|spk$k$|>}); text prompts are encoded by a
% per-row splitter parses the prompt on the first cue marker and dispatches it:
% \begin{itemize}\itemsep2pt
%   \item \emph{Caption and enrollment} $\rightarrow$ concat/self-attention.
%   \item \emph{Transcript} and \emph{speaker id} $\rightarrow$ cross-attention keys/values
%         \emph{plus} the audio-dimension grounding add. A single shared grounding block
%         serves both, and a multi-cue (transcript$+$speaker) row is routed to the word
%         objective.
%   \item \emph{Enrollment} $\rightarrow$ the reference clip is DAC-encoded and prepended
%         to the latent sequence (consumed only when enrollment audio is present;
%         text-only rows leave the slot masked).
% \end{itemize}
% adaLN pooling of the text uses separate projections for the caption path and the
% grounded (transcript/speaker) path to avoid weight reuse across cue types.

\vspace{-0.3cm}
\subsection{Alignment Module}
\vspace{-0.2cm}

The alignment module is designed to help the model capture the transcript and the temporal order in relation to the target speaker, comprising of a 2-layer bidirectional transformer that converts a sequence-level cue tokens into a frame-level target map. Let $H\in\mathbb{R}^{T\times d_{\text{mix}}}$ denote the WavLM representation of the mixture, where $T$ is the number of frames and $d_{mix}$ is the feature dimension; and let $E\in\mathbb{R}^{(L+1)\times d}$ denote the cue tokens (transcript or speaker id), where $L$ is the number of cue tokens, $d$ is their embedding dimension and the extra token is a learned \emph{null} token that represents the absence of a target. The module performs cross-attention with $H$ as queries and $E$ as keys and values. 

The output $O\in\mathbb{R}^{T\times d_v}$ is projected through an MLP and added to the diffusion latent.  
Intuitively, each mixture frame asks ``which cue token explains the target?'' and frames with no target attend to the null token.  The resulting map modulates the DiT separator toward the target time footprint.

\subsection{Auxiliary Alignment Loss}

We supervise the cross-attention map that links the cue tokens to the mixture frames. Let $\bar{A}\in{\mathbb{R}^{T\times(L+1)}}$ be the attention map averaged over heads.
Each row is normalized over keys (i.e., $\sum_{i} \bar{A}_{t,i}=1$ for every frame $t$). Let $v_i\in\{0,1\}$ select the valid (non-null) tokens and $tgt_t \in \{0,1\}$ marks target-active frames, obtained from the oracle target by RMS energy pooling. We define one loss for each cue routed through the alignment module. 

\noindent\textbf{Speaker Id}\; With tag mass $m_t=\sum_i v_i\,\bar{A}_{t,i}$, target frames should attend the tag and the rest the null (with $\beta$-weight on the null term):
\begin{align}
\label{eq:spk}
\resizebox{0.9\columnwidth}{!}{
$\mathcal{L}_{\text{spk}}=
-\frac{\sum_t \tgt_t\log m_t}{\sum_t \tgt_t}
-\beta\frac{\sum_t (1-\tgt_t)\log(1-m_t)}{\sum_t (1-\tgt_t)}$}
\end{align}

\noindent\textbf{Word-level Transcript}\; With text token spans $[f_0^i, f_1^i)$  from force-alignment, let  ${\footnotesize \phi_{t,i}=v_i\,\mathbf{1}[f_0^i\le t<f_1^i]}$, word mass
${\footnotesize \mu_t=\sum_i \bar{A}_{t,i}\,\phi_{t,i}}$, and ${\footnotesize h_t=\mathbf{1}[\sum_i\phi_{t,i}>0]}$
selects frames of a target word:
\begin{equation}
\label{eq:word}
\mathcal{L}_{\text{trans}
=\frac{\sum_t h_t\,(-\log \mu_t)}{\sum_t h_t}}
\end{equation}

\noindent\textbf{Total Objective}\; Combining the diffusion loss  $\mathcal{L}_{\text{mse}}$ with the task-specific alignment loss 
$\mathcal{L}_{\text{align}}\in\{\mathcal{L}_{\text{spk}},\mathcal{L}_{\text{wd}}\}$:
\begin{equation}
\mathcal{L}_{\text{total}} = \mathcal{L}_{\text{mse}}+\lambda\mathcal{L}_{\text{align}}
\end{equation}
In experiments, for loss functions above, we use $\beta$=0.5, $\lambda$=0.2.

\vspace{-0.3cm}
\input{table1}
\input{table2}
\input{table3}

\section{Experiments}
\vspace{-0.2cm}
% \vspace{-0.25cm}
\subsection{Training Data Construction and Hyper-parameters}
\label{sec:mixing}
\vspace{-0.2cm}

We train \method{} on synthesized mixtures from a multi-stem simulation pipeline. The source audio is a large internal collection of licensed data: $4.8$M speech, $1.1$M sound effects, $10$K music, and $2$K ambient noise samples. Each target-speech clip carries a word-level transcript and a caption primarily encoding gender and age. Each mixture is a $16$~s, $48$~kHz mono track with two or three speakers, simulated with room reverberation, ambient noise, sound effects, and music, arranged by time-layout templates covering common scenarios: 
\begin{itemize}[leftmargin=*,itemsep=1pt,parsep=0pt,topsep=1pt]
\item \textbf{partial overlap}: different speaker utterances begin at different onsets, with at least a one-second gap;
\item \textbf{turn-taking}: speakers switch in turns with ``chime-in" overlaps;
\item \textbf{full overlap}: speaker utterances share a common onset and offset, as traditional speech separation cases.
\end{itemize}
\method{} has roughly $0.8$B parameters and is fine-tuned from  \promptsep{}~\cite{wen2026promptsep} with AdamW~\cite{loshchilov2017decoupled} with a learning rate of $10^{-4}$. Denoising
follows a $v$-prediction objective. At inference, we use a DPM-Solver++(2M) SDE
sampler~\cite{lu2022dpmsolverpp} with a Karras schedule~\cite{karras2022edm} and $24$
steps.

\vspace{-0.25cm}
\subsection{Evaluation Datasets}
\vspace{-0.1cm}
We construct four evaluation sets from public and internal sources,  covering a range of tasks according to the cue labels each provides. Every set contains $5$k mixtures per task, which per-speaker enumeration expands into $10$k extraction instances, or $15$k for the three-speaker cases. \textbf{All evaluation sources are strictly disjoint from our training sources, and are thus fully out-of-domain}.

\vspace{0.15cm}
\noindent\textbf{CapSpeech-Mix}\; Built from CapSpeech~\cite{wang2026capspeech}. To emphasize distinctive speaker traits, we primarily retain age and gender in captions. 

\vspace{0.15cm}
\noindent\textbf{Libri2Mix}\; The $16$~kHz two-speaker benchmark~\cite{cosentino2020librimix,panayotov2015librispeech}, evaluated on both its
\emph{clean} and \emph{noisy} test sets. Only transcript and enrollment audio cues are available, so only these are evaluated.

\vspace{0.15cm}
\noindent\textbf{LibriTTS-R-Mix}\; Built from LibriTTS-R~\cite{koizumi2023librittsr}
test set. It carries no voice-attribute
descriptions, so the caption cue is not evaluated.

\vspace{0.15cm}
\noindent\textbf{Internal set}\; Built from an internal English speech corpus, disjoint from training sources. \textbf{This is a distict collection rather than a held-out split. }

\vspace{-0.25cm}
\begin{figure}[t]
% shift left so the heat-map's left edge (not the transcript labels) sits on
% the column margin; the labels overflow into the page margin.  Scale is set so
% the colour-bar tick labels land exactly on the right margin.
\noindent\makebox[\linewidth][l]{\hspace*{-0.1083\linewidth}%
\includegraphics[width=1.1116\linewidth]{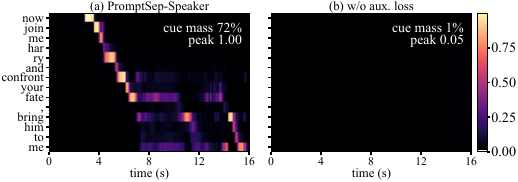}}
\vspace{-0.7cm}
\caption{Alignment-module attention (layer 2) between transcript tokens and mixture frames
for one two-speaker mixture, shared color scale, with (a) and without (b) the auxiliary
loss.}
\label{fig:attention}
\vspace{-0.5cm}
\end{figure}

\subsection{Baselines}
\label{sec:baselines}
\vspace{-0.1cm}
We compare \method{} against six baselines whose cues overlap with ours and as much as possible. Several concurrent TSE baselines that share a cue overlap with us are excluded because they do not release checkpoints: LLM-TSE~\cite{hao2023llmtse}, StyleTSE~\cite{huo2025styletse}, and two descriptor-based models~\cite{dai2025relative,dai2026twostage}. WeSep~\cite{zhang2026wesep} is a special case: the only checkpoint we coud locate supports the enrollment audio cue~\cite{realtse2026}. The included baselines are as follows: 

\vspace{0.15cm}
\noindent\textbf{DAE-TSE}~\cite{li2026daetse}\; The most recent keyword-guided BSRNN-based~\cite{luo2023music} TSE system, and to our knowledge the only released checkpoint is cued by transcript. Although it is trained mostly on 2-6 transcript words, it accepts transcripts of any length. 

\vspace{0.15cm}
\noindent\textbf{WeSep}~\cite{wang2024wesep}\; A BSRNN-based~\cite{luo2023music} TSE system on the
enrollment audio cue. We used its released checkpoint, and strip silence from the enrollment to match its training condition. 

\vspace{0.15cm}
\noindent\textbf{SoloSpeech}~\cite{wang2024solospeech}\; A state-of-the-art generative TSE
system on the enrollment audio cue. We used its released checkpoint. 

\vspace{0.15cm}
\noindent\textbf{\promptsep{}}~\cite{wen2026promptsep}\; Our prior caption-based and vocal-imitation-based generative source separator, without speaker-oriented training. It originally targets universal sound-event separation. 

\vspace{0.15cm}
\noindent\textbf{FlowSep}~\cite{yin2024flowsep} A caption-based generative source separator based on rectified-flow~\cite{liu2023rectifiedflow}.

\vspace{0.15cm}
\noindent\textbf{SAM-Audio}~\cite{samaudio}\; A caption-based and vision-based generative source separator. We used SAM-Audio-large for caption evaluation.

\vspace{-0.3cm}
\subsection{Objective Metrics}
\vspace{-0.1cm}
We report the scale-invariant signal-to-distortion ratio improvement
$\sisdri{}$~\cite{vincent2006bss,leroux2019sisdr}, a mel-spectrogram $L_2$ distance \cite{kumar2023dac}, and the DNS-MOS~\cite{reddy2022dnsmos}.
We further report F1 Decision Error~\cite{wen2026promptsep} to measure the model ability
to identify the correct temporal regions of target.
Audios are resampled to match each model setting during evaluation.

\vspace{-0.25cm}
\subsection{Subjective Listening Test}
\label{sec:subjective}
\vspace{-0.1cm}
We conducted a subjective listening test of 700
items, 100 for each of the four tasks plus three ablated task configurations. Samples are drawn
from the internal set as it is the only out-of-domain set \textbf{for all systems}. Every item presents a mixture as context, a ground-truth target as reference, and one system separation output. The subjects answer three MOS-style questions on a
five-point scale:

\begin{itemize}[leftmargin=*,itemsep=1pt,parsep=0pt,topsep=1pt]
\item \textbf{speaker accuracy}: how much of the separation is the reference speaker? 5 for completely target speaker and 0 for no target. 
\item \textbf{speech quality}: how is the audio quality of separation?
\item \textbf{content consistency}: does the output carry the target speaker words instead of the other? 
\end{itemize}
We collected 615 submissions per task (total 4,515) by 380 subjects.

\vspace{-0.3cm}
\section{Results}
\label{sec:results}
\vspace{-0.2cm}

\noindent\textbf{Transcript}\; Table~\ref{tab:main} evaluates four tasks by cues on two-speaker mixtures. We compare \method{} with DAE-TSE under the transcript cue. On the three out-of-domain sets, \method{} improves $\sisdri{}$ by $6.6$--$11.3$~dB over DAE-TSE, and attains much lower mel-$L_2$ and higher target-selection accuracy measured by F1 error. DAE-TSE leads only on Libri2Mix, as its in-domain benchmark.  Notably, DAE-TSE is a mask-based model, which tends to favor signal-level metrics than our generative formulation. Even so, \method{} remains mostly superior in signal-level metrics and consistently improves target-selection accuray across all sets. The human evaluation in Table~\ref{tab:subjective_eval_all} further confirms its advantage in both perceptual quality and separation accuracy,

\vspace{0.1cm}
\noindent\textbf{Enrollment Audio}\; On the CapSpeech, LibriTTS-R, and internal sets, \method{} outperforms SoloSpeech across $\sisdri{}$, mel-$L_2$, and F1. WeSep attains $\sisdri{}$ competitive with our model and higher than SoloSpeech, but this advantage does not carry over to other metrics. Similarly, Libri2Mix is in-domain  for SoloSpeech but out-of-domain for WeSep and \method{}. The subjective results further reveal that $\sisdri{}$ does not fully reflect perceptual quality particularly between a mask-based model and generative model: SoloSpeech and WeSep obtain very similar speaker accuracy and speech quality scores despite widely diverging $\sisdri{}$. Our \method{}, by contrast, remains consistent across objective and subjective evaluations on the enrollment audio cue.

\vspace{0.1cm}
\noindent\textbf{Speaker Id}\;  No baselines supports the tremporal-order speaker id cue, as~\cite{dai2025relative,dai2026twostage} release neither code nor checkpoints, so Table~\ref{tab:main} reports \method{} alone. Performance under this cue is on par with that under the other cues: relative to the full transcript, $\sisdri{}$ drops by only $0.41$, $0.81$, and $1.10$~dB on CapSpeech-Mix, LibriTTS-R-Mix, and the internal set respectively, while F1 is identical ($0.93$, $0.94$, $0.95$) and DNSMOS differs by at most $0.07$. This shows that, for two-speaker mixtures, \method{} achieves reliable target selection from an ordinal cue alone (i.e., speaker id) without any acoustic and lexical description of the speaker.

\vspace{0.1cm}
\noindent\textbf{Caption}\; \method{} outperforms all three caption-based baselines across both evaluation sets and every metric, with gains at least $12$~dB in $\sisdri{}$ and $0.17$ in F1 over the strongest baseline on the internal set. Both sets are out-of-domain for all models. \promptsep{} performs particularly poorly, with $\sisdri{}$ of ${-}32$ to ${-}37$~dB, F1 around $0.35$, and the lowest DNSMOS; qualitatively, we found it often extracts sound effects when given a speaker-specific cue, indicating that it fails to associate the cue with the target speaker. FlowSep behaves similarly ($\sisdri{}$ of ${-}29.3$ and ${-}31.7$~dB), and we found its outputs often stay close to the mixture rather than isolating the target. These two cases, however, yield more to a directional limitation than to the advantage of our model: as universal separators not trained substantially on speaker data, both baselines struggle with speaker extraction by design and data scarcity. The mitigation is SAM-Audio, which is trained on multi-speaker data. Though it separates the speaker far more reliably than other two baselines, it trails \method{} on every metric, which we attribute to its limited exposure to overlapping-speech data and conditions.

\vspace{0.1cm}
\noindent\textbf{Task Ablations}\; Table.~\ref{tab:stress} ablates the task configurations for transcript and speaker id conditions, degrading each cue and averaging the results over the three evaluations sets that support all of these cues. A partial transcript, giving fewer than $60\%$ of the full transcript words and further restricted between $3$ to $8$ words, yields $6.57$~dB mean $\sisdri{}$ with $0.89$ mean F1. This indicates that a few words are sufficient  to identify and track the target speaker but still leave a performance gap. The full-overlap transcript data constantly mix two speaker content from beginning to the end, where either of two speakers is never audible on its own. Our model yields $6.10$~dB  mean \sisdri{} with unchanged F1 to original transcript set. Each costs about $4$--$5$~dB of $\sisdri{}$ relative to the original transcript, but the high F1 shows the target speaker is still found in both cases. Our subjective results on Table~\ref{tab:subjective_eval_all} furthur confirms that the model performs within an acceptable gap among partial-transcript, full-overlap transcript and the original transcript sets. Such stress tests provide more insights on how the model conducts the trade-off on separation decision, quality and speaker tracking capability. Another stress test is to add a third speaker and this costs the most: the performance drops to $4.72$~dB \sisdri{} with F1 at $0.83$, a large decrease from 2-speaker cases. We attribute this degradation to both the naturally high difficulty of separation, and the WavLM representation: speaker id carries only order information, and thus WavLM must represent three voices within a single mixture latent, making the separation increasingly difficult as the number of speakers grows.

\vspace{0.1cm}
\noindent\textbf{Model ablations}\;
We consider two variants: one without the auxiliary alignment loss and one with the cross-attention direction reversed from frame-to-cue to cue-to-frame (i.e., audio attends to text or text attends to audio). Both degrade substantially in $\sisdri{}$ and F1 in Table~\ref{tab:stress}, while DNSMOS remains relatively strong, indicating that the ablations produce reasonable speech but fail to select the target reliably. Fig.~\ref{fig:attention} explains this behavior. With the auxiliary loss, transcript tokens attend to their corresponding speech frames, correctly localizing the target speaker's two utterances around the interferer. Without it, only $1\%$ of attention mass falls on the condition tokens, indicating that the model largely ignores the cue.

\vspace{0.1cm}
\noindent\textbf{Subjective listening test}\;
Table~\ref{tab:subjective_eval_all} confirms the objective results in human evaluation, with consistently high absolute scores. \method{} exceeds $4.0$ on $16$ of $21$ cells and on speaker accuracy, quality and content consistency for six of the seven tasks, including both transcript variants, which remain within $0.3$ MOS of the unstressed cue. Enrollment achieves $4.52$ and $4.53$ on these two criteria, while transcript achieves $4.39$ and $4.44$, respectively, on the five-point scale. In contrast, neither the baselines nor the ablations exceeds $4.0$ on any criterion, except WeSep, which reaches $4.23$ in content consistency but remains below \method{}. Overall, the listening test shows that \method{} not only separates speech with high perceptual quality, but also reliably identifies the intended speaker across different query modalities.

\vspace{-0.4cm}
\section{Conclusion}
\vspace{-0.3cm}
We presented \method{}, a generative latent-diffusion separation model that unifies four interchangeable cues, speaker id, transcript, enrollment, and caption. We introduced a new auxiliary alignment loss to efficiently condition speech representation. Across four evaluation sets and a listening test, \method{} outperforms specialized TSE systems and universal separation baselines. 
% Ablations show that the alignment loss is critical: without it, the model produces clean speech but largely selects targets at random. Notably, even a single ordinal token matches a full transcript in F1, although performance degrades with three speakers, which we leave for future work.

\vspace{-0.35cm}
\bibliographystyle{IEEEbib-compact}
\bibliography{references}

\end{document}

%% file: table1.tex
%% generated by make_table1.py -- do not edit by hand.
%% Sources: tab_main.tex and tab_libri2mix_compact.tex. Regenerate with:
%%     python make_table1.py
\providecommand{\abl}[1]{\textcolor{gray}{#1}}
\begin{table*}[t]
\centering
\footnotesize
\setlength{\tabcolsep}{2.5pt}
\vspace{-0.3cm}
\caption{Evaluation results. CapSp.: CapSpeech-Mix; L2Mix: Libri2Mix, averaged over its \emph{clean} and \emph{noisy} conditions; LTTS-R: LibriTTS-R-Mix;
Internal: a separate internal corpus, unused in training.
A dash marks a cue the corpus does not support.}
\label{tab:main}
\vspace{0.15cm}
\resizebox{\textwidth}{!}{%
\begin{tabular}{ll cccc cccc cccc cccc}
\toprule
& &
\multicolumn{4}{c}{SI-SDRi $\uparrow$} &
\multicolumn{4}{c}{mel-$L_2$ $\downarrow$} &
\multicolumn{4}{c}{DNSMOS $\uparrow$} &
\multicolumn{4}{c}{F1 Decision Error $\uparrow$} \\
\cmidrule(lr){3-6}\cmidrule(lr){7-10}\cmidrule(lr){11-14}\cmidrule(lr){15-18}
System & Cue & CapSp. & L2Mix & LTTS-R & Internal & CapSp. & L2Mix & LTTS-R & Internal & CapSp. & L2Mix & LTTS-R & Internal & CapSp. & L2Mix & LTTS-R & Internal \\
\midrule
DAE-TSE & transcript
& 2.13 & \textbf{11.13} & 2.41 & 1.21 & 41.70 & 18.50 & 39.30 & 41.90 & 2.37 & 2.61 & 2.67 & 2.68 & 0.69 & 0.82 & 0.68 & 0.62 \\
\textbf{Ours} & transcript
& \textbf{8.73} & 6.15 & \textbf{11.41} & \textbf{12.47} & \textbf{3.73} & \textbf{5.41} & \textbf{1.90} & \textbf{1.88} & \textbf{2.74} & \textbf{3.21} & \textbf{2.93} & \textbf{2.89} & \textbf{0.93} & \textbf{0.93} & \textbf{0.94} & \textbf{0.95} \\
\midrule
SoloSpeech & enroll
& 1.70 & \textbf{11.59} & 6.59 & 8.47 & 24.93 & 6.89 & 19.15 & 16.85 & \textbf{2.80} & \textbf{3.26} & \textbf{2.86} & 2.82 & 0.76 & \textbf{0.93} & 0.80 & 0.84 \\
WeSep & enroll
& \textbf{7.07} & 11.14 & 11.19 & 13.18 & 31.20 & 20.00 & 26.80 & 26.60 & 2.27 & 2.59 & 2.49 & 2.50 & 0.76 & 0.82 & 0.74 & 0.75 \\
\textbf{Ours} & enroll
& 4.01 & 5.04 & \textbf{11.79} & \textbf{13.30} & \textbf{7.48} & \textbf{6.00} & \textbf{2.31} & \textbf{1.84} & 2.69 & 3.21 & \textbf{2.86} & \textbf{2.89} & \textbf{0.87} & 0.92 & \textbf{0.95} & \textbf{0.96} \\
\midrule
\textbf{Ours} & spk id
& 8.32 & -- & 10.60 & 11.37 & 3.76 & -- & 2.90 & 2.51 & 2.67 & -- & 2.86 & 2.87 & 0.93 & -- & 0.94 & 0.95 \\
\midrule
PromptSep & caption
& -31.88 & -- & -- & -37.10 & 28.92 & -- & -- & 22.70 & 2.09 & -- & -- & 2.19 & 0.36 & -- & -- & 0.32 \\
FlowSep & caption
& -31.73 & -- & -- & -29.30 & 45.89 & -- & -- & 39.68 & 2.48 & -- & -- & 2.83 & 0.65 & -- & -- & 0.63 \\
SAM-Audio & caption
& -8.73 & -- & -- & -12.80 & 16.59 & -- & -- & 14.72 & 2.64 & -- & -- & 2.75 & 0.70 & -- & -- & 0.65 \\
\textbf{Ours} & caption
& \textbf{3.64} & -- & -- & \textbf{4.99} & \textbf{7.88} & -- & -- & \textbf{6.22} & \textbf{2.72} & -- & -- & \textbf{2.93} & \textbf{0.87} & -- & -- & \textbf{0.87} \\
\bottomrule
\end{tabular}}
\vspace{-0.35cm}
\end{table*}

%% file: table2.tex
%% generated by make_table2.py -- do not edit by hand.
%% Source: tab_main.tex. Regenerate with:  python make_table2.py
\begin{table}[t]
\centering
\footnotesize
\setlength{\tabcolsep}{4pt}
\vspace{-0.3cm}
\caption{Ablation studies across tasks with results averaged over CapSpeech, LTTS-R and internal sets. Task variants include using partial / full-overlap transcripts and three speakers. Model ablations include \emph{na} (without alignment loss) and \emph{td} (reverses the alignment cross-attention from frame-to-cue to cue-to-frame).}
\label{tab:stress}
\vspace{0.15cm}
\resizebox{\linewidth}{!}{%
\begin{tabular}{ll cccc}
\toprule
System & Cue & SI-SDRi $\uparrow$ & mel-$L_2$ $\downarrow$ & DNSMOS $\uparrow$ & F1 Dec. Err. $\uparrow$ \\
\midrule
Ours & transcript & 10.87 & 2.50 & 2.85 & 0.94 \\
Ours & transcript (partial) & 6.57 & 5.85 & 2.87 & 0.89 \\
Ours & transcript (full ov.) & 6.10 & 2.07 & 2.85 & 0.94 \\
\midrule
Ours & spk id (2 spk) & 10.10 & 3.06 & 2.80 & 0.94 \\
Ours & spk id (3 spk) & 4.72 & 7.47 & 2.85 & 0.83 \\
\midrule
Ours\,(na) & transcript & -21.43 & 20.18 & 2.79 & 0.56 \\
Ours\,(td) & transcript & -14.21 & 14.45 & 2.49 & 0.58 \\
Ours\,(na) & spk id (2 spk) & -15.40 & 17.87 & 2.80 & 0.62 \\
Ours\,(td) & spk id (2 spk) & -22.29 & 20.79 & 2.78 & 0.54 \\
\bottomrule
\end{tabular}}
\vspace{-0.35cm}
\end{table}

%% file: table3.tex
%% generated by make_table3.py -- do not edit by hand.
%% Source: tab_sbj_all.tex. Regenerate with:  python make_table3.py
\providecommand{\abl}[1]{\textcolor{gray}{#1}}
\begin{table}[t]
\centering
\footnotesize
\setlength{\tabcolsep}{4pt}
\vspace{-0.3cm}
\caption{Listening test on our internal set; MOS $\pm$ $95\%$ CI reported.}
\label{tab:subjective_eval_all}
\vspace{0.15cm}
\resizebox{\linewidth}{!}{%
\begin{tabular}{ll ccc}
\toprule
System & Cue & Speaker Accuracy & Speech Quality & Content Consistency \\
\midrule
DAE-TSE & transcript & $3.28 \pm 0.10$ & $3.55 \pm 0.11$ & $3.96 \pm 0.11$ \\
\textbf{Ours} & transcript & $\mathbf{4.39 \pm 0.12}$ & $\mathbf{4.16 \pm 0.11}$ & $\mathbf{4.44 \pm 0.12}$ \\
\midrule
DAE-TSE & trans.\,full-ov. & $3.09 \pm 0.12$ & $3.14 \pm 0.12$ & $3.67 \pm 0.14$ \\
\textbf{Ours} & trans.\,full-ov. & $\mathbf{4.19 \pm 0.15}$ & $\mathbf{4.01 \pm 0.12}$ & $\mathbf{4.18 \pm 0.15}$ \\
\midrule
DAE-TSE & trans.\,partial & $3.09 \pm 0.10$ & $3.41 \pm 0.11$ & $3.94 \pm 0.11$ \\
\textbf{Ours} & trans.\,partial & $\mathbf{4.20 \pm 0.15}$ & $\mathbf{4.04 \pm 0.12}$ & $\mathbf{4.17 \pm 0.15}$ \\
\midrule
SoloSpeech & enroll & $3.76 \pm 0.15$ & $3.63 \pm 0.12$ & $3.93 \pm 0.15$ \\
WeSep & enroll & $3.78 \pm 0.12$ & $3.63 \pm 0.13$ & $4.23 \pm 0.11$ \\
\textbf{Ours} & enroll & $\mathbf{4.52 \pm 0.10}$ & $\mathbf{4.07 \pm 0.11}$ & $\mathbf{4.53 \pm 0.10}$ \\
\midrule
\textbf{Ours} & spk id (2 spk) & $4.37 \pm 0.12$ & $3.73 \pm 0.13$ & $4.33 \pm 0.12$ \\
\textbf{Ours} & spk id (3 spk) & $3.78 \pm 0.17$ & $3.65 \pm 0.14$ & $3.69 \pm 0.17$ \\
\midrule
PromptSep & caption & $1.99 \pm 0.17$ & $3.03 \pm 0.17$ & $1.99 \pm 0.17$ \\
FlowSep & caption & $3.06 \pm 0.12$ & $3.09 \pm 0.12$ & $3.22 \pm 0.13$ \\
SAM-Audio & caption & $3.14 \pm 0.19$ & $3.74 \pm 0.14$ & $3.24 \pm 0.19$ \\
\textbf{Ours} & caption & $\mathbf{4.08 \pm 0.16}$ & $\mathbf{3.89 \pm 0.12}$ & $\mathbf{4.06 \pm 0.16}$ \\
\bottomrule
\end{tabular}}
\vspace{-0.35cm}
\end{table}